\documentclass[pra,showpacs,preprint,a4paper]{revtex4}

\usepackage{amssymb}
\usepackage{graphicx}
\usepackage{bm}

\usepackage{epsfig}
\usepackage{amsfonts}
\usepackage{color}

\begin{document}

\title{Spectral properties of a hybrid-qubit model based on a two-dimensional
quantum dot}
\author{Alba Y. Ramos}
\email{aramos@famaf.unc.edu.ar}
\author{Omar Osenda}
\email{osenda@famaf.unc.edu.ar}

\affiliation{Facultad de Matem\'atica, Astronom\'{\i}a y F\'{\i}sica,
Universidad Nacional de C\'ordoba, }
\affiliation{Instituto de F\'isica Enrique Gaviola, CONICET-UNC, Av. 
Medina Allende s/n, Ciudad Universitaria, X5000HUA C\'ordoba, Argentina }

\begin{abstract}
The design and study of hybrid qubits is driven by their ability  to
get along the best of charge qubits and of spin qubits, {\em i.e.} the speed of
operation of the former and the very slow decoherence rates of the latter ones.
There are several proposals to implement hybrid qubits, this works focuses on
the spectral properties of an one-electron hybrid qubit. By design, the
information would be stored in the electronic spin and the switching  between
the qubit basis states would be achieved using an external ac electric field.
The electron is confined in a two-dimensional quantum dot, whose confining
potential is given by a quartic potential, features that are typical of GaAS
quantum dots. Besides the confining potential
that characterizes the quantum dot there are two static magnetic fields applied
to the system, one is a large constant Zeeman field and the other one
has a constant gradient. We
study the spectral properties of the model Hamiltonian, a Scr\"odinger-Pauli
Hamiltonian with realistic parameters, using the Ritz method. In particular, we
look for regions of the parameter space where the lowest eigenenergies and
their eigenfunctions allow to define a qubit which is stable under perturbations
to the design parameters. We put special attention to the constraints that
the
design imposes over the magnetic fields, the tuning of the energy gap between
the qubit states and the expectation value of the spin operator where the
information would be stored. 
\end{abstract}
\date{\today}

\maketitle

\section{Introduction}

Semiconductors quantum bits are thought to be one of the most fertile fields 
to implement Quantum Information Processing and Computation
\cite{Hanson2007,Henneberger2008}. 
The advantages 
have been said countless times, from the incredible sophisticated techniques 
related to semiconductor technologies - such as integration \cite{Kawakami2014},
lithography\cite{Veldhorst2014} , 
ultra-pure sample preparation \cite{Veldhorst2014,Muhonen2014} and scalability
\cite{Kane1998}- to the 
peculiarities 
of the 
different kinds of qubits that are offered in semiconductors such as charge and 
spin qubits, exciton qubits and hybrid ones \cite{Henneberger2008}.

The disadvantages are, of course, a bit less heralded but, nevertheless, they 
are well known. One of the leading disadvantages lies in the fact that it is 
quite difficult to isolate a truly microscopic system (one electron, its spin, 
or the spin state of few electrons) from the semiconductor matrix in which it 
is embedded. The unavoidable interaction between the microscopic system 
chosen to carry the qubit and the environment that surrounds the systems leads 
to the loss of the quantum state coherence. Fortunately, the main mechanisms 
that produce decoherence are well known so it is possible to design {\em ad 
hoc} strategies to overcome or palliate their effects.

The term charge qubits was coined to design a qubit where the 
information is 
stored in the spatial degrees of freedom of one (or several) electron(s) 
\cite{Tanamoto2000,Ferron2013}. The 
main mechanism of decoherence, in this case, is owed to the coupling between 
the electron and the thermal phonons present in the semiconductor. For an
experimental study of charge relaxation in Si/SiGe DQD see the work by Wang
{\em et al.} \cite{Wang2013}. A similar study was performed by Srinivasa {\em
et al.} \cite{Srinivasa2013}, in their case they studied the
simultaneous spin-charge relaxation in DQD made of GaAs. 
As the 
information is stored in the spatial degrees of freedom, the quantum gates 
acting over the qubit take advantage of the strong coupling between electric 
(or electromagnetic) fields and the electron charge. The strength of the 
coupling is determined by the spatial extent of the electron wave function, 
this fact led, for instance,  to the development of the exciton qubit
\cite{Troiani2000,Biolatti2000,Bonadeo1998} where the coupling is smaller than
in the electronic case. An 
exciton, the electron-hole pair produced in a semiconductor when a valence band 
electron is promoted to the conduction band, also is coupled to the thermal 
phonons but the coupling with the phonons depend on the {\em difference} 
between the electron and hole states \cite{Haug2004}. So, using charge qubits
based on states of particles with opposite charge
provides a fairly 
simple 
mechanism to attenuate the effects of the phonon-induced decoherence
allowing 
picoseconds gate operation times compared with the characteristic time 
scale  of the decoherence, around the nanoseconds \cite{Henneberger2008} . 

The implementations where the 
quantum information is stored in the intrinsic angular momentum of one
\cite{Loss1998} or more electrons are called spin qubits (for the implementation
using two electrons
see \cite{Levy2002,Holleitner2002}, and for three electrons
\cite{Gaudreau2006,Korkusinski2007}). The decoherence mechanism is 
owed to the 
coupling 
between the electron spin and the nuclear spin of the atoms in the 
semiconductor. The  maturity and development of this 
particular implementation is heavily indebted to the techniques developed in the 
area of Nuclear Magnetic Resonance (NMR), from the pulse sequence techniques 
\cite{Witzel2007,Shi2013,Burkard2002} up to 
dynamic noise suppression techniques \cite{Petta2005,Taylor2007}. 
The double-well quantum dot 
proposal is a 
direct application of the NMR ideas to the implementation
of spin qubits 
\cite{Petta2005}. Some 
quite spectacular advances have been achieved in dealing with the decoherence 
mechanisms using qubits based on quantum dots embedded in ultra-pure Silicon 
samples very recently \cite{Morello2010, Xiao2010, Maune2012}. Despite that it 
is possible to obtain very 
long 
coherence times, spin qubits are not (yet) an  accomplished answer to the
problems of  integration and the interaction between other qubits 
because their design has become more and more involved. Of particular concern
are the long operation times 
inherent to the weak coupling between the magnetic dipole moment of the 
qubit quantum state and a magnetic field applied to it. 

Hybrid qubits \cite{Tokura2006,Shi2012,Jing2014}, those in which the information
is stored in the electron spin 
but the gate operation is provided by electric fields coupled to the electron 
charge, seem a good compromise between the two ways of approaching the problem 
described above. In particular, some years ago, Tokura {\em
et al.} 
\cite{Tokura2006} proposed a 
hybrid qubit, 
that besides long coherence and short operation times would take advantage of 
the very low strength of the noise spectrum when the energy gap between the 
qubit states lies in the $\mu\mbox{eV}$ range. The design proposed is elegant 
and sophisticated, employing a nano-wire to confine one electron in one 
dimension, a double well potential in this dimension an a pair of magnetic 
fields, one applied along the direction of the nano-wire and the other 
perpendicular to it. The proposal was aimed to implement single electron spin
resonance (SESR) in quantum dots.

Despite some inherent peculiarities of the design in Tokura {\em et al.}
\cite{Tokura2006}, the hybrid qubit model has several features that are general
enough to make a closer look at it very compelling.
In this work, we study the spectral properties of a hybrid qubit inspired
by the one proposed in \cite{Tokura2006} but easing some confining
requirements, in particular the model considered in this work is not one
dimensional so the orbital angular momentum plays a non-trivial role.
The model
hybrid qubit and its Hamiltonian, a two-dimensional Schr\"odinger-Pauli
equation, is presented and analyzed in some detail in Section~\ref{sec:model}.
The numerical method employed to obtain a highly accurate spectrum and
eigenfunctions is described in Section~\ref{sec:numerical-method}. Since the
model depends on several parameters, a detailed analysis of the lowest lying
eigenvalues, their eigenfunctions and the expectation values that characterize
the possibility of using the system as a  qubit is presented in
Section~\ref{sec:results}, paying some attention to the stability of the system
when its parameters are changed. As it will be shown, a thorough analysis of
the model eigenvalues and eigenfunctions is a bit arduous but necessary to
study the dynamics of the system when an external driving is applied. So,
despite that it is our goal to study the performance of the system under time
dependent external forcing, we deferred this investigation for the sake of
conciseness.
Finally, a discussion about the results and its
implications is the subject of Section~\ref{sec:conclusions}.

\section{Model and Hamiltonian}\label{sec:model}

Tokura and collaborators \cite{Tokura2006} proposed an hybrid qubit based on a 
very particular quantum dot. Actually, their proposal relied heavily on the 
ability to confine the electron in one dimension using a nano-wire. Besides,  
along  the direction of the nano-wire, the electron would be confined in a 
double well potential. To define the qubit basis states two magnetic fields 
were needed, one of them along the nano-wire, and the other perpendicular to 
it. Following the work of Tokura we denote the direction along the nano-wire as 
the $\hat{z}$ direction. The field along the $\hat{z}$ direction is constant in 
time and spatially uniform, while the perpendicular one has a constant gradient 
along the coordinate $z$. Again, following Tokura, the perpendicular magnetic 
field points in the $\hat{x}$ direction.

The one-dimensional confinement had a 
twofold purpose, on one hand it allowed to neglect the contributions owed to the 
other spatial coordinates and consequently the Hamiltonian considered by Tokura 
{\em et al.} did not included any term related to the orbital angular momentum. 
On the other hand, despite the spatial dependence of one of the magnetic fields 
applied to the quantum dot, the spin angular momentum terms could be treated 
easily and the qubit basis states were always eigen-states of the spin angular 
momentum in the $\hat{x}$ direction. Of course, in the case that the 
confinement is not strictly one-dimensional, this can not be taken for granted. 
The swapping between the qubit basis state was achieved applying an electric 
external driving. 

In this work we analyze a very similar system, that allows to analyze a number 
of features that a one-dimensional model can not. Besides,  our model does not 
impose such severe constraints to a possible actual implementation.

\begin{figure}[floatfix]
\begin{center}
\includegraphics[scale=0.4]{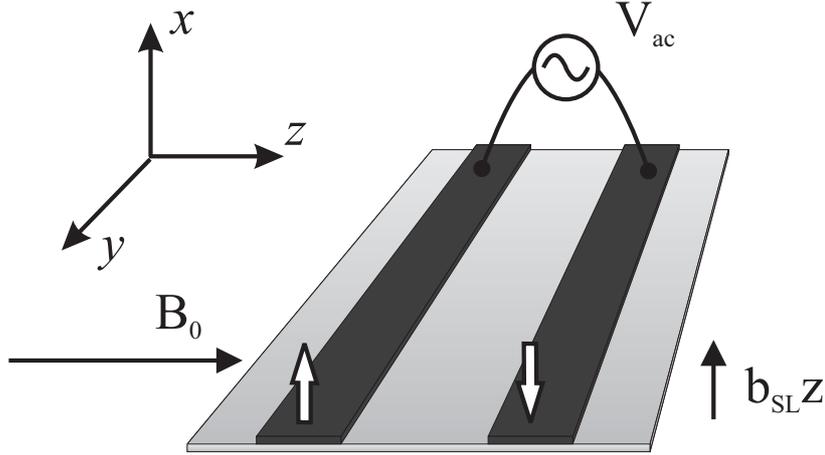}
\end{center}
\caption{\label{graf:qd} A cartoon showing the geometry of the nano-structure
considered. The darker grey  stripes in the light-grey surface point to where
the double potential well is located. The two external magnetic fields, $B_0$
in the $\hat{z}$ direction and $b_{SL} z$ in the $\hat{x}$ direction are shown.
The white
arrows show the direction in which the spin would point if the system works
properly. The electric potential $V_{ac}$ is used to produce the switching
of the electron from one potential well to the other. }
\end{figure}

Figure~\ref{graf:qd} shows a cartoon depicting the geometry, coordinate system 
and applied fields to the  two-dimensional quantum dot that is under scrutiny 
in this work. The surface represents the two-dimensional region where the 
electron is bound. Along the $\hat{z}$ direction an external static slanting
magnetic 
field $B_0$ is applied, also in this direction, the electron is confined by
a double well potential, 
in this case depicted schematically by the two black stripes that lie parallel 
to the $\hat{y}$ direction. We do not consider the presence of a confining 
potential in the $\hat{y}$ direction anyway, as we will show, the magnetic 
field $B_0$ effectively bound the electron in this direction. In any case, this 
assumption has the same physical implications that to consider that the 
characteristic length of the confinement in the  $\hat{y}$ direction is larger 
than the radius of the first Landau levels associated to a magnetic field of 
strength $B_0$. The other magnetic field that is applied to the system has 
a constant gradient, $b_{SL}$, here we follow Tokura {\em et al.}
\cite{Tokura2006} that considered a magnetic field of the form
$(b_{SL}z,0,b_{SL}x)$ 

Finally, the electron is forced to jump between the two potential wells by the 
electric driving provided by the potential $V_{ac}$. 

Despite that the system described above is two-dimensional, the main physical 
behavior should be similar to the behavior observed in the system proposed by 
Tokura {\em et al}, {\em i.e} when the electron is in a given potential well 
the electron magnetic dipole moment should point in the direction that 
minimizes the energy, upwards in the rightmost potential well and downwards in 
the leftmost potential well.  Of course, in a two-dimensional system this is 
{\bf not} equivalent to say that when the electron is ``located'' in one given 
potential well it will be in the corresponding eigenstate of the Pauli 
operator $\sigma_x$, which is proportional to the spin angular momentum of the 
electron. 

This last statement remark the elegance of the proposal made by 
Tokura {\em et al}, since the one dimensional system grants that the eigenstates
of 
the Hamiltonian are also eigenstates of $\sigma_x$, simplifying the analysis of 
the spectrum, the eigenstates and their properties.

The electron Hamiltonian, $\mathcal{H}$, can be written as
\begin{eqnarray} \label{eq:hamiltonian-general}
 {\cal 
H}=\frac{1}{2\,m^{\star}}\left[\vec{\mathbf{\sigma}}\cdot\left(-i\hbar\vec{
\mathbf{
\nabla}}-\frac{e}{c}\vec{\mathbf{A}}(\mathbf{r},t)\right)\right]^2+V(\mathbf{r})
,
\end{eqnarray}
where $\mathbf{A}$ is the vector potential, $\sigma$ are the Pauli matrices, 
$V(\mathbf{r})$ the double-well potential and $m^{\star}$ is the electron 
effective 
mass. Along this work we use the GaAs effective mass, so $m^{\star}/m = 0.041$.

Since the total magnetic field applied to the QD is given by
\begin{equation}\label{eq:magnetic-field}
 \mathbf{B}=(b_{SL}z,0,B_0+b_{SL}x),
\end{equation}
then, calling $\mathbf{B_0} = (0,0,B_0)$ and 
$\mathbf{B_1} = (b_{SL}z,0,b_{SL}x)$, we get that
\begin{eqnarray}\label{eq:vector-potential}
 \mathbf{A}_1&=&(0,\frac{b_{SL}}{2}(x^2-z^2),0)\nonumber\\
 \mathbf{A}_0&=&-\frac{B_0}{2}(y,-x,0) ,
 \end{eqnarray}
using the symmetrical gauge for the vector potential.

Replacing the expressions obtained for the vector potentials and magnetic 
fields, Equations~\ref{eq:magnetic-field},\ref{eq:vector-potential}, in the 
Hamiltonian, Equation~\ref{eq:hamiltonian-general}, and assuming a confining 
potential
\begin{equation}\label{eq:double-well}
V(\mathbf{r}) = V(z) = \frac{m^{\star} \omega_0^2 }{8 b^2} (z^2 -a^2)^2 -\gamma
\frac{\hbar \omega_0}{a} z,
\end{equation}
it can be found that
\begin{eqnarray}\label{eq:hamiltonian-v1}
 {\cal H}&=&-\frac{\hbar^2}{2\,m^{\star}} \nabla^2
+V(z) + \frac{i\hbar\,e}{m^{\star}
c}\left[-\frac{B_0}{2}\frac{\partial}{\partial 
x}+\left(\frac{b_{SL}}{2}(x^2-z^2)+\frac{B_0}{2}x\right)\frac{\partial}{
\partial 
y}\right]+\nonumber\\
 &+&\frac{e^2}{2m^{\star} 
c^2}\left[\frac{B_0^2y^2}{4}+\frac{b_{SL}^2}{4}(x^2-z^2)^2+\frac{b_{SL}B_0}{2}
(x^2-z^2)x+\frac{B_0^2}{4}x^2\right]\nonumber\\
&-&\frac{\hbar\,e}{2m^{\star}\,c}\,\left[b_{SL}\,z\,
\mathbf{\sigma}_x+\left(B_0+b_{SL}\,
x\right)\mathbf{\sigma}_z\right].
\end{eqnarray}

The confining potential, Equation~\ref{eq:double-well}, is a slight 
generalization of the well-known quartic double-well potential, that has two 
wells centered around $\pm a$, where $\omega_0$ sets the energy scale and the 
parameter $b$, that has length units, allows to change the height of the 
potential in $z=0$. The linear term in Equation~\ref{eq:double-well} plays a
double purpose, on one hand it allows the possibility that an electron
confined in it has a non-degenerate ground state and, on the other, allows to
study the quite possible scenario where both potential wells have different
depth. It is clear that the difference in depth between the potential wells in
their centers ($\pm a$) is given by $\Delta_{pw} = 2 \gamma \hbar \omega_0$. If
$\hbar \omega_0$ is in the order of $10 \, \mbox{meV}$ and $\gamma \approx
10^{-3}$, then $\Delta_{pw}$ is on the order of the energy gap between the
qubit basis states for which the device is designed for. We will return latter
to the effect of this term over the behavior of the system.
Finally, note that for $b=a$ and $a$ large enough, the ground state energy of 
the electron is a fraction of $\hbar \omega_0$.

The spectrum of the Hamiltonian in Equation~\ref{eq:hamiltonian-v1}, is quite 
complicated to calculate because its lacking of symmetries, and the coupling 
between the spin degrees of freedom. Moreover, there are several length and 
energy scales that characterize the physical behavior. It is interesting to 
note that, using matrix notation, the Hamiltonian \ref{eq:hamiltonian-v1} can 
be written as
\begin{eqnarray} \label{eq:hamiltonian-matrix}
\mathcal{H}& =&
\left(\begin{array}{cc} H&0\\ 0&H 
\end{array}\right)+H_1\otimes\sigma_x+H_2\otimes\sigma_z\nonumber\\
&=&\left(\begin{array}{cc} H&0\\ 0&H 
\end{array}\right)+\left(\begin{array}{cc} 0&H_1\\ H_1&0 
\end{array}\right)+\left(\begin{array}{cc} H_2&0\\ 0&-H_2 
\end{array}\right)\nonumber\\
&=&\left(\begin{array}{cc} H+H_2&H_1\\ H_1&H-H_2 
\end{array}\right)
\end{eqnarray}.

So, neglecting the terms related to the $x$ coordinate and introducing scaled 
variables $y^{\prime} = y/a$, $z^{\prime} = z/a$, we get
\begin{eqnarray}\label{eq:two-dimensional-hamiltonian}
  \frac{{\cal 
H}_{2d}}{\hbar\omega_0}&=&-\frac{(\hbar\,\omega_a)}{2\,(\hbar\,\omega_0)}\left[
\frac{\partial^2}{\partial {z'}^2}+\frac{\partial^2}{\partial 
{y'}^2}\right]+\frac{(\hbar\,\omega_0)}{8\,(\hbar\,\omega_a)}({z'}
^2-1)^2 -\gamma z' + \nonumber\\
&-&\frac{i\,\hbar\omega_c^*}{\hbar\omega_0}\frac{b_{SL}\,a}{2\,B_0}{z'}^2\frac{
\partial}{\partial 
y'}+\frac{\left(\hbar\,\omega_c^*\right)^2}{2\,(\hbar\omega_a)(\hbar\omega_0)}
\left[\frac{{y'}^2}{4}+\left(\frac{b_{SL}\,a}{2\,B_0}\right)^2{z'}^4\right]
\nonumber\\
&-&\frac{\hbar\,\omega_c^*}{2\,\hbar\omega_0}\left[\frac{b_{SL}\,a\,z'}{B_0}
\sigma_x+\sigma_z\right],
\end{eqnarray}
where $\omega_c^{\star} = \frac{e B_0}{\hbar m^{\star} c}$ is the Larmor 
frequency 
associated to the magnetic field $B_0$,  $\omega_a = \frac{\hbar}{m^{\star} 
a^2}$, and the subscript $2d$ stands for ``two-dimensional''. 
The frequency $\omega_a$ is introduced to make evident that 
$\frac{{\cal 
H}_{2d}}{\hbar\omega_0}$ is effectively dimensionless, which is also true for 
the ratios $b_{SL}a/B_0$, and $\omega_c^{\star}/\omega_0$.

Obtaining accurate numerical approximations to the spectrum and eigenstates of 
the Hamiltonian \ref{eq:two-dimensional-hamiltonian} is the subject of the next 
Sections, but before it is worth to pay some more attention to the 
analysis of the Hamiltonian \ref{eq:two-dimensional-hamiltonian}. 

The different terms in the Hamiltonian are characterized by ratios between four 
characteristics energies: $\hbar \omega_0$, $\hbar \omega_c^{\star}$, $\hbar 
\omega_a$ and $\hbar \omega_{SL}$. The latter frequency is the Larmor frequency
of the 
magnetic field $b_{SL} a$. Often, the role of $\omega_a$ is 
better understood recalling that $\frac{\hbar \omega_0}{\hbar \omega_a} = 
\left(\frac{a}{\ell_0}\right)^2$, where $2a$ is the distance between the 
quartic potential minima, and $\ell_0$ is the characteristic length of a 
quantum  harmonic oscillator with ground state energy $\hbar \omega_0/2$. If 
the electron is well localized in a given potential well, then $\ell_0 < a$ and 
$\frac{\hbar \omega_0}{\hbar \omega_a} >1$. In this case when $a$ increases its 
value, the two wells of the quartic potential become more and more separated 
and the spectrum consists in a set of quasi-degenerate levels, at least for
small enough values of $\gamma$, which are the ones that we will consider in
this work. The  
situation described above is where the terms associated to both magnetic fields
are relevant to 
produce a pair of states suitable to be used as the qubit basis states. In this 
scenario the energies are ordered, $\hbar \omega_0 > \hbar \omega_a >\hbar 
\omega_c^{\star} \sim \hbar \omega_{SL}$. 

\section{The numerical method}\label{sec:numerical-method}

The  Ritz variational method \cite{Merzbacher} has been used to obtain high
accuracy 
approximations to the spectrum and eigenstates of many different problems
\cite{Osenda2007,Ferron2009,Ramos2014}. 
A sensible one-particle basis set must be chosen to 
apply the method. From the 
analysis described in the 
Section~\ref{sec:model}, it is reasonable to pick a basis set, 
$\lbrace \chi_{\lbrace i\rbrace}\rbrace_{i=1}^M$, such that
\begin{equation}\label{eq:basis-set}
\chi_{\lbrace i\rbrace} =  \Psi^{p}_{k,n}  \alpha ,
\end{equation}
where $k=1,\ldots,L$,  $n=1,\ldots,N$, $p =\pm$, $\alpha$ is any of the
eigenstates of 
$\sigma_z$, $\lbrace i \rbrace \equiv (k,n,p,\alpha)$  and
\begin{equation}\label{eq:spatial-basis}
\Psi^{\pm}_{k,n}(y,z) = \phi_k(y) \psi_n^{\pm}(z) . 
\end{equation}
The functions in Equation~\ref{eq:spatial-basis} are given by
\begin{equation}\label{eq:basis-oscillator-y}
 \phi_k(y) = \mathcal{N}_k H_k(\mu y) e^{-\mu^2 y^2/2},
\end{equation}
\begin{equation}\label{eq:basis-z}
\psi_n^{\pm}(z) = C_n^{\pm} \left(\psi_n^{+a}(z) \pm \psi_n^{-a}(z) \right) ,
\end{equation}
with 
\begin{equation}\label{eq:basis-oscillator-z}
\psi_n^{\pm a}(z) = \mathcal{N}_n H_n(\eta z) e^{-\eta^2 (z\mp a)^2/2},
\end{equation}
where $H_m$ are the Hermite polynomials, and $\mathcal{N}_m$ is a normalization
constant, as well as $C^{\pm}_n$.
The functions in Equation~\ref{eq:basis-z} are chosen as a combination of
functions centered in each potential well following \cite{Tokura2014}.
 It is clear that both kind of functions, $\phi_k$ and $\psi_n^{\pm a}$ in
Equations~\ref{eq:basis-oscillator-y} and \ref{eq:basis-oscillator-z} are 
harmonic oscillator-like eigenfunctions where the non-linear variational 
parameters $\eta$ and $\mu$ are to be chosen to minimize the actual value of 
the approximate ground state energy obtained when  the Ritz method is applied. 
It is worth to mention that $L$ and $N$ is the number of functions $\phi$'s and 
$\psi^{\pm}$'s included in the basis set, so the the basis set has 
$M=4\ldotp L \ldotp N$ dimension.

The different length scales already mentioned in the preceding Section suggest 
the election of a basis set with functions well localized in both potential 
wells of the quartic potential, a condition satisfied by the functions in 
Equation~\ref{eq:basis-z}. For $a$ large enough the non-linear variational 
parameter $\eta$ should be, roughly speaking, of the same order than 
$1/\ell_0$. While the functions $\phi_k$ are centered around zero, the 
functions $\psi_n^{\pm a}$ are centered around $\pm a$ so, chosen in this way, 
the functions $\psi_n^{\pm}(z)$ are able to take into account two length 
scales, $a$ and $\ell_0$. 

The variational eigen-functions, $\Xi_j^v$, can be obtained from the solutions 
of the generalized eigenvalue problem
\begin{equation}\label{eq:eigenvalue-problem}
\mathbf{H} \mathbf{c}_j = E_j^v \, \mathbf{S} \, \mathbf{c}_j ,
\end{equation}
where the elements of the matrix $\mathbf{H}$  are given by the 
matrix elements of the two-dimensional Hamiltonian, 
Equation~\ref{eq:two-dimensional-hamiltonian}
\begin{equation}\label{eq:matrix-elements}
\mathbf{H}_{ij} = \left\langle \chi_i | \mathcal{H}_{2d}| \chi_j \right\rangle
,
\end{equation}
and the overlap matrix elements are given by
\begin{equation}
\mathbf{S}_{ij} = \left\langle \chi_i |\chi_j \right\rangle .
\end{equation}
Once the variational eigenvalues $E_j^v$ and  eigenvectors $\mathbf{c}_j$ are 
obtained, the approximate eigenfunctions can be written as
\begin{equation}
\Xi_j^v = \sum_i \mathbf{c}_j^i \chi_i ,
\end{equation}
where $\mathbf{c}_j^i$ is the $i$-th element of the column vector 
$\mathbf{c}_j$.

Numerous features  characterize the 
numerical procedure and the calculation of intermediate quantities necessary to 
carry it out.
The generalized eigenvalue problem, Equation~\ref{eq:eigenvalue-problem}, can 
be tackled using the well-known Cholesky method. The matrix elements on 
Equation~\ref{eq:matrix-elements} can be obtained exactly using the algebraic 
methods associated to creation and annihilation operators that are usually 
employed when dealing with the quantum harmonic oscillator or related problems, 
since all the terms in the Hamiltonian, 
Equation~\ref{eq:two-dimensional-hamiltonian}, are given in terms of powers or 
derivatives of both coordinates, $y$ and $z$.  The 
analytical expressions are evaluated using MP Fortran algorithms and the 
generalized eigenvalue problem is solved using an accuracy of at least sixteen 
significant figures.

\begin{figure}[floatfix]
\begin{center}
\includegraphics[width=7cm]{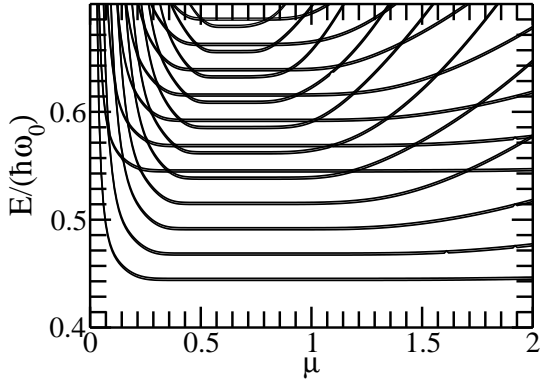}
\end{center}
\caption{\label{graf:stabilization} The lowest lying numerical eigenvalues
obtained for Hamiltonian~\ref{eq:two-dimensional-hamiltonian} using the Ritz
method {\em vs} the non-linear variational parameter $\mu$. The spectrum is
formed by near degenerate pairs of eigenvalues. The eigenvalues are stable for
an interval, whose width is larger for smaller eigenvalues. The actual size of
the interval can be extended by using increasingly larger basis sets. }
\end{figure}

One of the main concerns when the Ritz method is applied has to do with the 
{\em stability} of the eigenvalues calculated, {\em i.e.} the method is 
unstable if the minimum of the ground state energy depends  on the 
particular values chosen for the non-linear variational parameters. A sensible 
base, with a large enough number of basis functions, should provide stable 
eigenvalues for  relatively large intervals of of the non-linear variational 
parameters. Figure~\ref{graf:stabilization} shows the approximate spectrum 
calculated using the Ritz method as a function of the parameter $\mu$, for a QD 
with $\hbar \omega_0=30 \mbox{meV}$, $a=30$nm, $\gamma=-10^{-3}$, $B_0=0.5
\mbox{T}$, $b_{SL}a=2$T, and a 
symmetric base with $L=20$ and $N=20$, $\eta=4$. 

From the Figure~\ref{graf:stabilization} it is clear that the numerical method 
is able to provide more than thirty stabilized eigenvalues for $\mu \in 
(0.5,0.75)$, in particular, for $\mu\in (0.5, 1)$ the ground state energy is 
constant with a relative error smaller that $2\times 10^{-5}$. Of course, if 
the parameters of the system are changed too much it is necessary to analyze 
the behavior of the eigenvalues as functions of the non-linear variational 
parameters again, which can be a bit cumbersome. Anyway, once the values of 
$\eta$ and $\mu$ are fixed, the matrix elements on 
Equation~\ref{eq:matrix-elements} must be calculated only once to analyze 
different sets of QD parameters.

\section{Results}\label{sec:results}

\subsection{The spectrum of the quartic potential}

The quartic potential, Equation~\ref{eq:double-well} without the linear term,
has been used numerous times to model the low energy spectrum of quantum dots.
As we are interested in the study of the quantum dot spectrum described by
the Hamiltonian in Equation~\ref{eq:two-dimensional-hamiltonian} in the regime
where the dominant (or at least bigger) contribution comes from the QD binding
potential, it makes sense to look at the low energy part of the spectrum of an
one-dimensional Hamiltonian given by
\begin{equation}
 {\cal H}_{1d}=-\frac{\hbar^2}{2\,m^{\star}}\left[\frac{\partial^2}{\partial 
z^2}\right]+\frac{m^{\star}\,\omega_0^2}{8\,a^2}(z^2-a^2)^2,
\end{equation}
in particular, we look for regions of the $(\hbar \omega_0, a)$ space where the
energy gap between the ground state and the first excited one is on the order
of tens of micro or just micro electron volts ($\mu eV$).

\begin{figure}[floatfix]
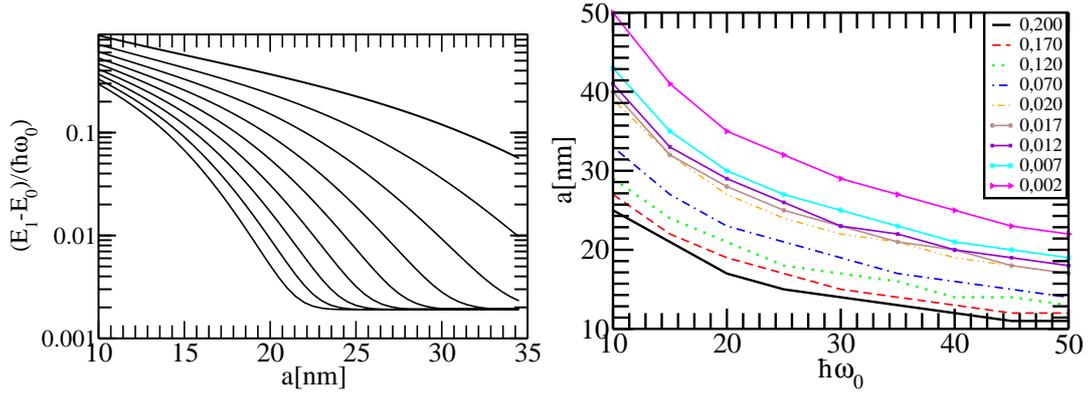

\begin{center}
\includegraphics[width=7cm]{fig3_a.eps}
\includegraphics[width=7cm]{fig3_b.eps}
\end{center}
\caption{\label{graf:gap} a) The scaled energy gap {\em vs} the potential well
distance for different values of $\omega_0$. From top to bottom the lines
correspond to $\hbar \omega_0 = 10$ meV, $\hbar \omega_0 = 15$ meV, and so on
up to $\hbar \omega_0 = 50$ meV. b) The contour lines of the scaled gap in the
$(\hbar \omega_0,a)$ plane. From top to bottom the different contour lines
correspond to different values of $\Delta$.}
\end{figure}

Figure~\ref{graf:gap} a) shows the behavior of the energy gap scaled with the
factor $\hbar \omega_0$, for different strengths of $\hbar\omega_0$. The curves 
show that the scaled energy gap as a
function of $a$ displays three regions where its behavior is different. For
small enough values of $a$ the energy gap decays accordingly with the quartic
potential, for intermediate values  the energy gap decays exponentially, as it
is expected for any problem with two potential wells that are being separated,
and for large values of $a$ the energy gap becomes fairly independent of $a$.
The asymmetry between the potential wells is responsible for this last feature.

From  Figure~\ref{graf:gap} a),
it is clear that for a large enough separation $a$ it is possible to obtain a
gap on the order of $10^{-3} \hbar \omega_0$. Besides, with the data in
Figure~\ref{graf:gap} a) it is possible to
draw the contour plot of the scaled gap as shown in Figure~\ref{graf:gap} b),
choosing the minimum value of $a$ compatible with the corresponding value of the
energy gap.
Interestingly, the curves in panel b), that correspond to fixed values of the
scaled energy gap, can be accurately fitted with the function $a = A \times
(\hbar \omega_0)^{-\frac12}$, where $A$ is a constant that depends on the
actual value of $(E_1-E_0)/\hbar \omega_0$.

The data in Figure~\ref{graf:gap}
provides enough evidence that there is a scenario, that we call
``well-separated wells'', where the gap between the ground state and the first
excited one can be fairly well tuned to the desired values. For instance,
Tokura {\em et al.} considered that a gap in the order of the $\mu eV$ could be
useful to palliate the effects of the environmental noise over the information
stored in the electron spin . Obviously,  there is always a distance for
which the gap is as small as desired, that this distance is within the tens of
nano-meters reinforces the idea that the scenario can be obtained with
``realistic'' parameters. Note that, if the potential well are asymmetric, then
the minimum value of the energy gap will be on the order of the difference
between the potential well depths.

\subsection{A good qubit}

\begin{figure}[floatfix]
\begin{center}
\includegraphics[width=7cm]{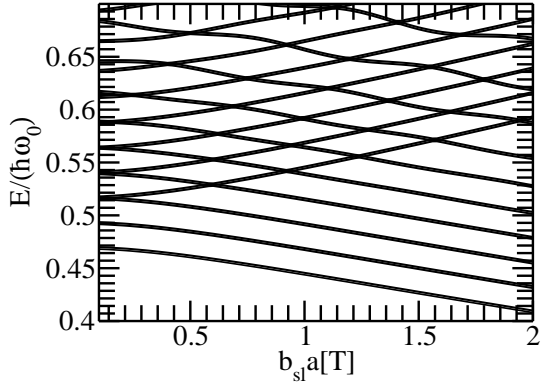}
\end{center}
\caption{\label{graf:spectrum} The approximate spectrum {\em vs} the 
magnetic field $b_{SL} a$. The spectrum mainly consists in pairs of near
degenerate
eigenvalues so, at the scale used in the Figure, each line is formed by two
very close energy levels. For the parameters used, see the text, the three
first pairs of levels do not show avoided crossings.}
\end{figure}

To begin with the study of the physical behavior of the QD system, it is useful 
to recall that our search is directed to found QD systems with a pair of states 
that can be used as the basis states of a hybrid qubit. In particular, since 
the information will be stored in the electron spin but, as the swapping 
between that pair of states would be done using an electric driving potential, 
the pair of states should have  spin expectation values easily distinguishable 
(ideally they will be two different eigenstates of $\sigma_x$) and be located 
at different wells of the quartic potential. This analysis suggests that the we 
must consider together the  behavior of the spectrum, and the expectation value 
of the $\sigma_x$ and $z$ operators.

Figure~\ref{graf:spectrum} shows the behavior of the variational spectrum as a 
function of the field $b_{SL}a$. The other parameters of the system are those 
used to obtain the data in Figure~\ref{graf:stabilization}, {\em i.e.}
$a=30$nm, $\hbar \omega_0 = 30$meV, $B_0=0.5$T, $\gamma=-10^{-3}$. 
All the eigenvalues shown are nearly degenerate pairs, clearly, this 
corresponds to the regime of ``well-separated potential wells''. It is
interesting to note that although both, the energy of the 
fundamental state and the energy of the first excited state, depend on the 
applied field and the difference between them remains basically constant with a 
value of $2\times 10^{-3} \, \hbar \omega_0$. So, the gap between these states 
belong to the $\mu \mbox{eV}$ domain, a region of energies where the noise
spectrum that affects the spin degrees of freedom is low enough 
to allow a coherent manipulation of the state \cite{Tokura2006}. Anyway, despite
all the apparent 
advantages that show these two states it is necessary to verify that they are 
sufficiently distinguishable when a measure of the component $x$ of the 
intrinsic angular momentum is performed. 

\begin{figure}[floatfix]
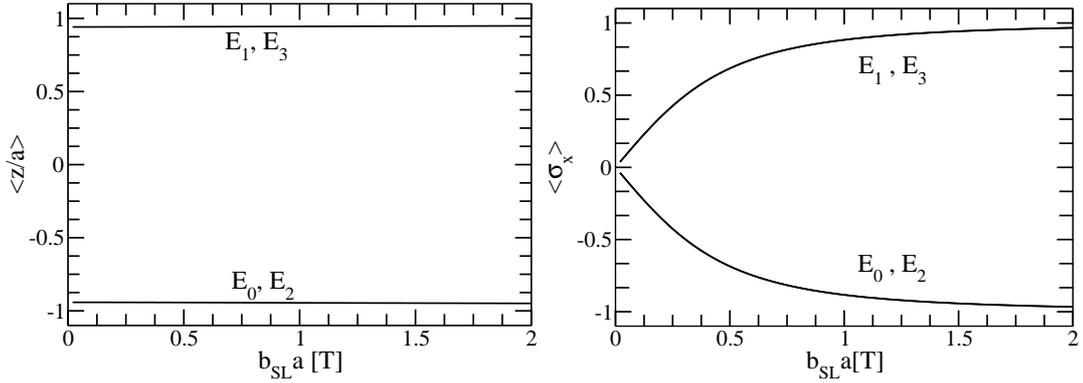

\begin{center}
\includegraphics[width=7cm]{fig5_a.eps}
\includegraphics[width=7cm]{fig5_b.eps}
\end{center}
\caption{\label{graf:expectation-values} The expectation value of operators
$z/a$
(panel a))  and $\sigma_x$ (panel b)) as a function of the magnetic field
$b_{SL} a$ for the first four eigenstates. In each panel, the labels $E_0, E_1,
E_2$ and $E_3$ correspond to the values obtained for the ground, first
excited, second and third excited states, respectively. It is clear that in
this
regime, the eigenstates are strongly localized around the centers of the
potential wells, the ground state is localized in the leftmost well, the first
excited state in the rightmost and so on.}
\end{figure}

The scenario described in the paragraph above 
can be better understood by looking at the behavior of the expectation values 
of the $\sigma_x$ and $z$ operators.
Figure~\ref{graf:expectation-values} shows 
these quantities for the first few eigenstates as functions of the field 
$b_{SL}a$. 

Figure~\ref{graf:expectation-values} a) shows the expectation value of the 
position operator $z$ as a function of the magnetic field $b_{SL}a$. This 
 Figure clearly shows the confirmation that in this 
regime the eigenstates are localized in just one  well of the quartic 
potential, and the fact that the ground state is localized in the leftmost 
well, the first excited state in the rightmost, the second in the leftmost and 
the third in the rightmost. Besides, from this last argument, we infer that the
behavior 
observed in the spectrum, the eigenvalues are monotone functions of the 
magnetic field unless the corresponding eigenstate jumps from one potential 
well to the other, as can be observed for large enough eigenvalues , see
Figure~\ref{graf:spectrum}. So, 
each avoided crossing observed in Figure~\ref{graf:spectrum} is produced when 
a state that is localized in one well jumps to the other. Finally, for the set 
of parameters under analysis here, the expectation value of the operator $z$ 
shows that the electron is very close to the minimum of the potential well 
where it is localized since $\langle z \rangle \approx a$. As we will show 
latter, this is not necessarily so in other regions of the parameter space.

Figure~\ref{graf:expectation-values} b) has some more information to give. It 
shows the behavior of the expectation value of $\sigma_x$ evaluated for the 
first four eigenstates and as a function of the magnetic field $b_{SL} a$.
Obviously, for 
zero magnetic field the expectation value is also zero. When the strength of 
the magnetic filed is increased the absolute value of the expectation value 
also grows, but it does not reach the unity, {\em i.e.} the  
Hamiltonian eigenstates never become eigenstates of $\sigma_x$. So the minimum
strength of the magnetic field that allow to have a reliable qubit depends 
on how well the spin measurement process can distinguish between the two lowest 
states. 

\subsection{Sensitivity to variations in the QD parameters}

So far, we have presented almost the best case scenario, {\em i.e.} a QD with a 
set of parameters that possesses a pair of states with all the desirable 
properties stated in the work by Tokura, an energy gap on the order of the 
$\mu\mbox{eV}$, both eigenstates are almost eigenstates of $\sigma_x$ for 
large enough transversal field, and all the parameters used are ``realistic'', 
at least in the sense that are compatible with actual QD. Anyway, to gain more 
physical insight we proceed to analyze how much those parameters can be changed 
without spoiling the good features of the qubit basis states.

\begin{figure}[floatfix]
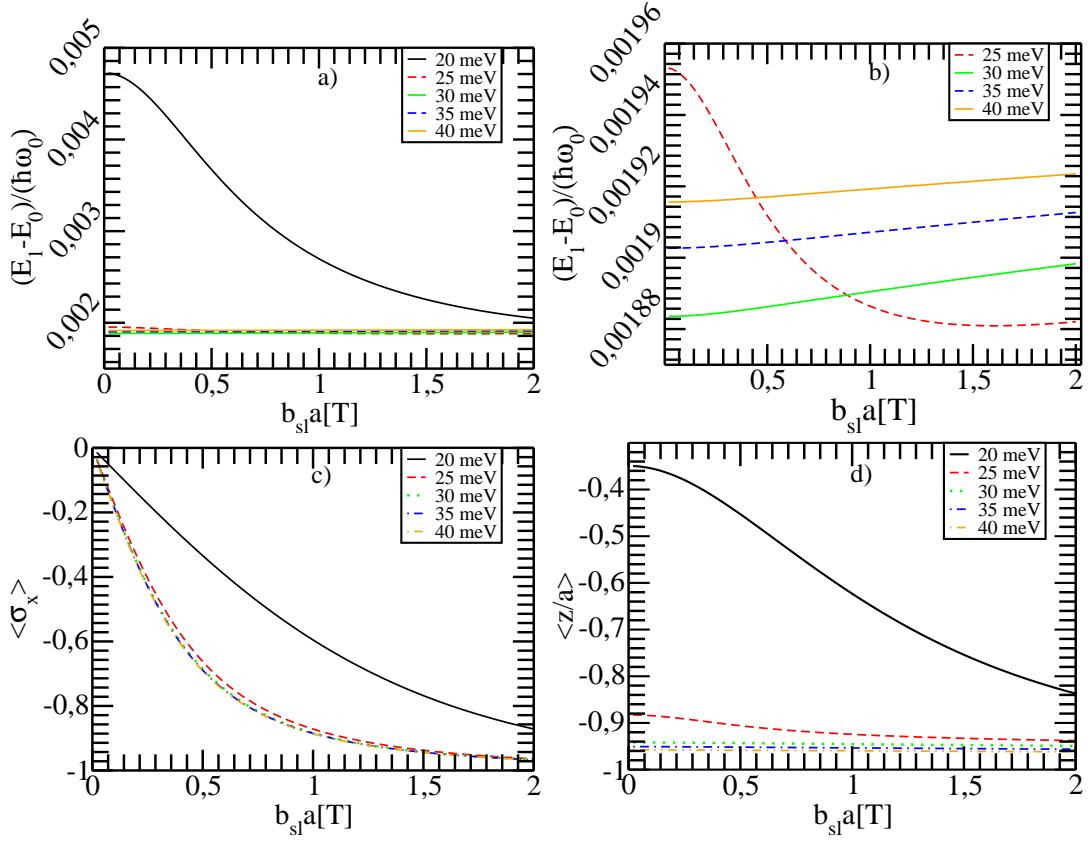

\begin{center}
\includegraphics[width=7cm]{fig6_a.eps}
\includegraphics[width=7cm]{fig6_b.eps}
\includegraphics[width=7cm]{fig6_c.eps}
\includegraphics[width=7cm]{fig6_d.eps}
\end{center}
\caption{\label{graf:stability-w0} a) and b) The scaled gap of energy between
the qubit basis states, c) and d) the ground state's expectation value  of the
$\sigma_x$ and $z$ operators, respectively, as functions of the field strength
$b_{SL} a$ and for different quantum dots characterized by energies
$\hbar\omega_0 = 20,25,30,35$ and $40$ meV.}
\end{figure}

Regrettably, the model QD has almost too many parameters to show here a 
comprehensive exploration of the parameter space. There are three parameters 
that define the quantum dot, the separation between the minima of the quartic 
potential $a$, the frequency $\omega_0$ and the height of the barrier between 
both wells that is regulated by $b$. As we will show, the net effect over the 
spectrum, the energy gap between the ground state and the first excited state, 
the expectation value of operators $z$ and $\sigma_x$ is pretty similar 
irrespective of which specific parameter is changed. 

Figure~\ref{graf:stability-w0} compiles the results obtained changing the value 
of the frequency $\omega_0$, in particular a) shows the behavior of the energy 
gap between the ground state and the first excited one {\em vs} the magnetic 
field for five different values of $\omega_0$, b)  shows a close up of  
panel a) (because the quite different scale of the data shown in a)), c) shows 
the expectation value of $\sigma_x$ and d) shows the expectation value of $z$. 

Roughly speaking, the behavior of the different quantities can be understood 
noting that for a given separation of the potential wells, in our case $2a$, 
there are values of $\omega_0$ that take the quantum dot outside the regime of
``sufficiently separated wells'', leading to small values for the expectation 
values of $z$ (in absolute value, see the top curve in panel d)) and $\sigma_x$
(see the top 
curve in panel c)). Interestingly, increasing the strength  of the magnetic 
field restores the system to the sufficiently separated wells,  but the 
restoring effect requires increasingly larger values of the magnetic field as
the 
value $\omega_0$ decreases. Anyway, for a design value of $\hbar \omega_0=30 
\mbox{meV}$ the system is pretty stable since there will not be appreciable 
changes in the behavior of the states associated to the qubit even for 
deviations of $\omega_0$ as large as  $20\%$.

In some sense, the argument above is consistent with the results sketched in
Figure~\ref{graf:gap} b), {\em i.e.} changing the parameters of the QD along a
contour-line in the $(\hbar \omega_0,a)$ plane produces very much the same
quality of qubit basis states or, in other terms for a given $a$ there is
always a ``weak-enough'' value of $\hbar \omega_0$ that would result in a pair
of states well-suited to be used as qubit basis states. This is clearly to be
expected, but Figure~\ref{graf:gap} and Figure~\ref{graf:stability-w0} provide
the quantitative evidence that this scenario can be achieved for realistic
parameters. Anyway, since the two lowest eigenstates of the Hamiltonian are not
eigenstates of the $\sigma_x$ operator, the ability to distinguish between
states when a measurement of $\sigma_x$ is performed imposes a certain
constraint on the minimum value that can take the field gradient $b_{SL}$.

Changing the other two parameters that define the confining potential for the
electron, $b$ and $\gamma$, has pretty much the same effect over the spectrum
and the expectation values that changing $\hbar \omega_0$ or $a$,  loosely
speaking it is fair to say that if the gap between the ground and the first
excited state becomes too large because $b$ becomes smaller and smaller then
it is always possible to separate the wells making $a$ larger and larger in
such a way to keep the energy gap between the desired limits. On the other
hand, for larger values of the asymmetry factor $\gamma$ it is easier to obtain
larger values for the expectation values of $\sigma_x$, anyway, it is not
always easy to design asymmetric double quantum dots and the energy gap
between the qubit basis state can become very large leaving rapidly the domain
around the $\mu \mbox{eV}$.

The large Zeeman magnetic field $B_0$ is used in NMR to
set the energy scale between the different polarization states of the magnetic
nuclei in the sample under scrutiny. Larger magnetic fields imply a stronger
signal since the net polarization, at a given temperature, depends on the
different populations of the ``up'' and ``down'' levels. The nuclei are heavy
enough to ignore what happens with the orbital degrees of freedom, but this is
not the case with electrons. The proposal of Tokura {\em et al.}
\cite{Tokura2006} was intended to provide a qubit where single electron spin
resonance could be performed. As we will discuss next, what is good for NMR it
is not necessarily so good for SESR in quantum dots.

\begin{figure}[floatfix]
\begin{center}
\includegraphics[width=7cm]{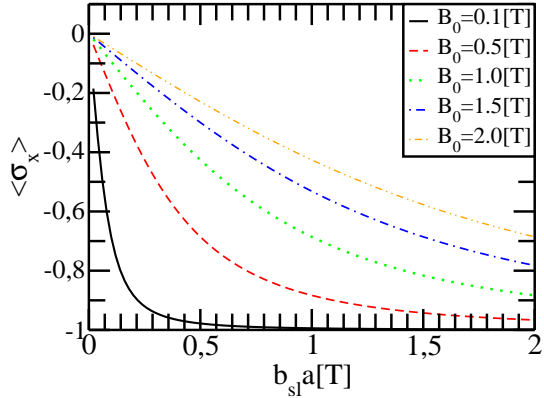}
\end{center}
\caption{\label{graf:varying-B0} The ground state expectation value of the
operator $\sigma_x$ {\em vs} the magnetic field $b_{SL}a$ for different
$B_0$ strengths. From top to bottom, the curves correspond to strengths of
$2$, $1.5$, $1$, $05$ and $0.1$ Tesla.}
\end{figure}

Figure~\ref{graf:varying-B0} shows the behavior of the ground state expectation
value of the $\sigma_x$ operator as a function of $b_{SL}$ and for different
strengths of $B_0$. The data show clearly that if the Zeeman field becomes
larger  it is then necessary to resort to increasingly large values of the
gradient in
order to obtain  large enough expectation values of $\sigma_x$. In some sense,
is in this Figure where the two-dimensional character is revealed more
plainly. Besides,
Figure~\ref{graf:varying-B0}  shows what was previously enunciated, {\em i.e.}
that higher values of
Zemann field  may worsen the qubit  behavior, diminishing the strength of
the
signal if a measurement of $\sigma_x$ is performed.

\section{Conclusions and Discussion}\label{sec:conclusions}

The results shown in this work do not preclude the possibility of implementing
a qubit suitable to study SESR, up to a point our results just strengthen some
of the constraints already present in the original proposal.
Nevertheless, the net effect of considering a two-dimensional system seems to
be the need of stronger field gradients $b_{SL}$ to achieve the same results
than in a one-dimensional system.

On the other hand, the design is pretty robust against perturbations to their
design parameters, even having into account the possible asymmetry between the
potential wells depth. So, there is a broad range of parameters where the qubit
can be implemented, but the controllability, stability and possible probability
leakage should be investigated carefully. Work around this lines is currently
in development.

 It would be
interesting to explore the transition from a two-dimensional system, like the
one studied in this work, to a one-dimensional one, reducing the confinement
length in the $\hat{y}$ direction from much larger than $2a$ to smaller than
$a$, but most probably other basis set for the Ritz method must be employed.




\acknowledgments
We would like to acknowledge  SECYT-UNC,  and CONICET 
for partial financial support of this project and to Federico Pont for critical
reading of the manuscript.

\end{document}